\documentclass[pra,10pt,twocolumn,reprint,superscriptaddress,floatfix,showpacs]{revtex4-2}

\usepackage[utf8]{inputenc}
\usepackage[T1]{fontenc}     
\usepackage[british]{babel}  
\usepackage[sc,osf]{mathpazo}
\usepackage[scaled=0.86]{berasans}  
\usepackage[colorlinks=true, citecolor=blue, urlcolor=blue]{hyperref}  
\usepackage{graphicx} 
\usepackage{subfig}
\usepackage[babel]{microtype}  
\usepackage{amsmath,amssymb,amsthm,bm,amsfonts,mathrsfs,bbm} 

\usepackage{xspace}  
\usepackage{pgfplots}
\usepackage{xcolor,colortbl}
\def\ba{\begin{equation}}
\def\ea{\end{equation}}
\def\bea{\begin{eqnarray}}
\def\eea{\end{eqnarray}}
\def\ben{\begin{equation*}}
\def\een{\end{equation*}}
\def\bean{\begin{eqnarray*}}
\def\eean{\end{eqnarray*}}
\def\bma{\begin{mathletters}}
\def\ema{\end{mathletters}}
\def\bi{\begin{itemize}}
\def\ei{\end{itemize}}

\newcommand{\be}{\begin{equation}}
\newcommand{\ee}{\end{equation}}

\newcommand{\kommentar}[1]{}

\newcommand{\forget}[1]{}

\begin{document}

\title{Detecting Nontrilocal Correlations In Triangle Networks}
\author{Kaushiki Mukherjee}
\email{kaushiki.wbes@gmail.com}
\affiliation{Department of Mathematics, Government Girls' General Degree College, Ekbalpore, Kolkata-700023, India.}

\begin{abstract}
Correlations in quantum networks with independent sources exhibit a completely novel form of nonclassicality in the sense that the nonlocality of such correlations can be demonstrated in fixed local input scenarios. Before the pioneering work by M.O.Renou, \emph{et al.,} in \cite{gis1}, the nonlocal feature of such network correlations was directly attributable to standard Bell nonlocality. In \cite{gis1}, the authors provided some of the first examples of triangle network correlations, whose nonlocality cannot be deduced from Bell-CHSH nonlocality. To date, a complete characterization of such scenarios is yet to be provided. Present work characterizes correlations arising due to fixed local measurements in a triangle network under a source independence assumptions. Precisely speaking, a set of criteria is framed in the form of Bell-type inequalities, each of which is necessarily satisfied by trilocal correlations. Possible quantum violation of at least one criterion from the set is analyzed, which in turn points out the utility of the set of criteria to
detect nonlocality (nontrilocality) in quantum triangle networks. Interestingly, measurement on a local product state basis turns out to be sufficient to generate nontrilocal correlations in some quantum networks. Noise tolerance of the detection criteria is discussed followed by a generalization of the framework for demonstrating correlations in any $n$-sided polygon where $n$ is finite.
\end{abstract}

\maketitle

	
\section{Introduction}\label{intro}
Violation of Bell's inequalities is considered a cornerstone in the study of quantum foundations\cite{Bel,Bel2}. It points out the impossibility of
interpreting quantum predictions in terms of any physical model dependent only on the local variables. Violations of such correlator-based inequalities
are referred to as Bell nonlocality and the corresponding experimental set-up is commonly known as the standard Bell experiment\cite{brunrev}. A standard Bell experiment involves a single source that distributes particles to two or more distant observers. In case the source distributes an entangled quantum state and the parties perform suitable local measurements, the resulting correlations may be Bell nonlocal in nature. \\
\par Over the years, extensive research activities have revealed multifaceted features of Bell nonlocality. However, in the last decade, the study of nonlocality has witnessed remarkable advancement beyond the paradigm of the Bell scenario. Such a trend of study is motivated to manifest behavior of quantum correlations associated with network topologies compatible with different practical tasks\cite{indr1,internet1,lee,internet2}$.\,$Apart from the presence of multiple distant parties, involvement of multiple independent sources is an inherent feature of such network scenarios\cite{BRA,BRAN,FRITZ}$.$ In a network scenario each source typically connects to a proper subset of parties in contrast to the standard Bell scenario. \\
\par $\textmd{Independence of sources in quantum networks adds}$ new physical insights in analyzing non classicality of network correlations\cite{gis1,BRA,BRAN,bilo1,km1,km2,bilo2,bilo3,km3,bilo4,km4,km5,bilo5,ejm,bilo6,bilo7}$.\,$For instance consider the simplest network of three parties, Alice ($A$), Bob ($B$) and Charlie ($C$), and two independent sources
$\mathcal{S}_1,\mathcal{S}_2.$ $\mathcal{S}_1$ and $\mathcal{S}_2$ distribute  particles among pairs (A,B) and (B,C) respectively (see Fig.\ref{fig}, for $n$$=$$2$). This network structure is referred to as the \textit{Bilocal Network}\cite{BRA,BRAN}.
In contrast to the standard tripartite Bell experiment, initially, they do not share any common past. Bob performs joint measurement, in particular, the Bell state measurement (BSM\cite{BRAN}) on two particles (received from $\mathcal{S}_1$ and $\mathcal{S}_2$), whereas Alice and Charlie each perform local measurements. Under suitable measurement choices, nonbilocal tripartite correlations may be
shared at the end of the experiment. \\
\par over time, the bilocal network has been generalized on basis of various factors such as increasing number of parties, sources\cite{km1,km3}, a pattern of arranging independent sources and parties\cite{bilo1}, etc. Quantum network structures with independent sources are compatible with repeater networks\cite{indr1,indr3}, where entanglement gets distributed via the entanglement swapping procedure among initially uncorrelated nodes.
In general, under the source independence constraint, a set of network local correlations being non convex\cite{BRAN}, framing Bell-type inequalities becomes a challenging task$.\,$Over the years different Bell-type network inequalities showing quantum violation have been established (see \cite{birev} for a review).  \\
\par In an $n$-local network ($n$ denoting the number of sources), quantum correlations can exhibit nonlocality (in sense of non $n$-locality) when some or all parties perform single measurements, i.e., they have fixed input in the sense that they do not choose from a collection of measurements\cite{BRAN,gis1}.
This contributes to another striking difference with a standard Bell experiment where each of the parties must choose an input randomly and
independently from a collection of two or more inputs\cite{brunrev}$.\,$Such a choice of inputs forms one of the key factors required for Bell nonlocal behavior of the corresponding correlations. The source independence assumption, referred to as the \textit{$n$-local} constraint, thus reduces requirements for demonstrating non classicality in quantum networks\cite{gis1,BRA,BRAN}.\\
Non $n$-locality even in absence of randomness in input selection points out a genuine form of nonlocality typical of network scenarios.
In \cite{gis1}, such a notion of nonlocality was referred to as \textit{nonlocality without inputs}. However, in literature, before the publication
of \cite{gis1}, all instances of network nonlocality in fixed input measurement scenarios relied on Bell nonlocality in the sense that violation of $n$-local inequality can be exploited in terms of violation of the standard Bell-CHSH inequality\cite{cla}. This in turn questioned whether
network correlations can truly exhibit any form of non-classicality that will be completely independent of standard Bell nonlocality. In \cite{gis1}, the authors first clarified the doubt. They provided examples of correlations that are non trilocal (non $n$-local for $n$$=$$3$) in a triangle network.
They also justified that such non-classical correlations are typical of measurement scenarios involving independent sources
and thereby can never be simulated in a standard Bell experiment. \\
\par In \cite{gis1}, the authors provided specific instances of non trilocal correlations. For that, they analyzed the structure of correlations generated in a triangle network where each source generates the same pure two-qubit state and each observer performs specific joint measurements\cite{gis1}.
In this context, it will be interesting to frame a criterion that can detect the nontrilocality of correlations (in triangle networks) in general.\\
\par This work characterizes the set of trilocal correlations arising in a triangle network where each party performs a fixed measurement with
four outputs. For such manifestation, a set of non-linear Bell-type inequalities has been constructed. This set emerges as a necessary condition for trilocality.
Violation of at least one of these inequalities suffices to ensure the nontrilocality of corresponding correlations$.\,$Under a suitable measurement context, there exist families of quantum states violating at least one such inequality. Interestingly, it is observed that some of these inequalities can detect nontrilocal quantum correlations even when none of the two-qubit states is Bell-CHSH nonlocal. Consequently, the set of inequalities emerges as the first trilocal criteria in a fixed input scenario whose violation cannot be traced back to the violation of the Bell-CHSH inequality. Besides, it is observed that nontrilocal correlations can be simulated in the network involving separable states and/or choosing a product state basis as a local measurement basis for each party. An inequality from the set is used for detecting pure entanglement. The procedure of constructing trilocal inequalities can be generalized beyond the triangle topology so that the network involves more than three independent sources.\\
\par The rest of the work is organized as follows$:\,$in sec.\ref{mot}, the motivation of the present discussion is illustrated. Sec.\ref{pr} provides some basic
pre-requisites.$.\,\,$A set of trilocal inequalities for triangle network is introduced in sec.\ref{main}. Sec.\ref{quan} deals with quantum violation of members from the set thereby characterizing quantum nontrilocal correlations in a triangle network. In sec.\ref{ent}, the utility of the set of trilocal criteria is discussed from the perspective of entanglement detection. The noise tolerance of some members from the set of the inequalities is analyzed in sec.\ref{noise}. In sec.\ref{comp}, the comparison is made between some of the trilocal inequalities for a triangle network with that of the existing trilocal inequality for a linear trilocal network\cite{km1}. The triangle network scenario is generalized in sec.\ref{poly}. Finally, the discussion ends with some concluding remarks in sec.\ref{conc}.
\section{Motivation}\label{mot}
Network scenarios with multiple independent sources correspond to different quantum network topologies\cite{internet1,lee,internet2,indr1,indr2,indr3}.
From experimental perspectives, quantum non $n$-locality has emerged as a non-classical resource in different network-based practical tasks\cite{birev}. Moreover, non-classicality (non $n$-locality) in fixed input scenario points out some intriguing features of network correlations\cite{gis1} which have no analog in standard Bell experiment.
Detection of such type of network nonlocality motivates the present discussion. Regarding the fixed input scenario, only a few particular
instances of quantum nontrilocal correlations exist in a triangle network\cite{gis1}. Recently, triangle networks along with other network scenarios have been analyzed from different perspectives\cite{nr1,nr2,nr3}. However, to the best of the author's knowledge, there does not exist any criterion to detect nontrilocality in such a scenario. In this context, a set of criteria in form of non-linear Bell-type inequalities for detecting nontrilocality
 will be derived here. Testing any such criterion is supposed to be experimentally feasible as it is based on correlators. \\
 \par A network is said to be open if at least one of the parties receives particles from a single source. Considering the parties as vertices, the arrangement of parties in such a network does not correspond to the vertices of any polygon. Non linear $n$-local inequality exists for open type $n$-local networks\cite{birev}. In this context, it becomes interesting to derive Bell-type inequalities for closed $n$-local networks. A closed network refers to the arrangement of the parties and the sources such that each pair of parties receives particles from a common source. So in a closed network, the parties can be interpreted as vertices of a polygon. The simplest form of closed network topology is considered here. The establishment of trilocal triangle network inequalities and then multi-directional analysis of their quantum violation forms the basis of the current discussion$.\,$Once framed, a comparison of triangle trilocal inequalities with that of existing trilocal inequality for a linear network\cite{km1} becomes essential. Besides, it is also crucial to generalize the present study for analyzing $n$-local closed networks.\\
 \par At this junction, it may be noted that Finner inequalities\cite{fin1} can be used to detect whether a given set of correlations is generated in a network or not\cite{fin2}. However, in a network, the corresponding Finner inequality is not only compatible with a local distribution but is also satisfied when the sources generate bipartite quantum states or arbitrary no signaling resources\cite{fin2}. So any Finner inequality can be used to differentiate between
 network (compatible with Finner inequality considered) correlations from those which are not generated in the network\cite{fin2}. But no such inequality can be used to exploit nonlocal behavior (if any) of the corresponding set of network correlations. \\
\par It may be noted that because of the source independence assumption, trilocal hidden variable models form a subclass of general local hidden variable (LHV) models. The inexplicability of network correlations by any trilocal hidden variable model does not ensure nonlocality in general as there may exist some more general LHV models explaining those correlations. So non trilocality forms a restricted form of nonlocality. Yet, a study of nontrilocality is important as this form of non-classicality is compatible with network topologies involving distant sources such that each source independently distributes particles to a subgroup of parties. Such networks are commonly used in the field of standard information technology (e.g. internet).

With progress in quantum information science, quantum counterparts of this type of network are forming building blocks in different information processing tasks\cite{nr4,internet1,nr6}. Apart from theoretical advancement, there has been rapid technological development towards scalable quantum networks which in turn leads to multiple uses of quantum networks\cite{indr1,internet1,internet2}. Hence, from a practical perspective, studying the non $n$-locality of quantum correlations warrants attention.
\section{Preliminaries}\label{pr}
\subsection{Bloch Matrix Representation of Two Qubit State}
Let $\rho$ denote any two-qubit state. The corresponding density matrix of $\rho$ is given in terms of Bloch parameters:
\begin{equation}\label{st4}
\small{\varrho}=\small{\frac{1}{4}(\mathbb{I}_{2}\times\mathbb{I}_2+\vec{\mathfrak{a}}.\vec{\sigma}\otimes \mathbb{I}_2+\mathbb{I}_2\otimes \vec{\mathfrak{b}}.\vec{\sigma}+\sum_{j_1,j_2=1}^{3}w_{j_1j_2}\sigma_{j_1}\otimes\sigma_{j_2})},
\end{equation}
with $\vec{\sigma}$$=$$(\sigma_1,\sigma_2,\sigma_3), $ $\sigma_{j_k}$ denoting Pauli operators along three mutually perpendicular directions ($j_k$$=$$1,2,3$). $\vec{\mathfrak{a}}$$=$$(x_1,x_2,x_3)$ and $\vec{\mathfrak{b}}$$=$$(y_1,y_2,y_3)$ denote
local bloch vectors ($\vec{\mathfrak{a}},\vec{\mathfrak{b}}$$\in$$\mathbb{R}^3$) corresponding to party $\mathcal{A}$ and $\mathcal{B}$ respectively with $|\vec{\mathfrak{a}}|,|\vec{\mathfrak{b}}|$$\leq$$1$ and $(w_{i,j})_{3\times3}$ stands for the correlation tensor matrix $\mathcal{W}$ (real matrix).
Components $w_{j_1j_2}$ of $\mathcal{W}$ are given by $w_{j_1j_2}$$=$$\textmd{Tr}[\rho\,\sigma_{j_1}\otimes\sigma_{j_2}].$ \\
$\mathcal{W}$ can be diagonalized by applying suitable local unitary operations\cite{gam,luo},where the simplified expression is then given by:
 \begin{equation}\label{st41}
\small{\varrho}^{'}=\small{\frac{1}{4}(\mathbb{I}_{2}\times\mathbb{I}_2+\vec{\mathfrak{m}}.\vec{\sigma}\otimes \mathbb{I}_2+\mathbb{I}_2\otimes \vec{\mathfrak{n}}.\vec{\sigma}+\sum_{j=1}^{3}\mathfrak{t}_{jj}\sigma_{j}\otimes\sigma_{j})},
\end{equation}
Correlation tensor in Eq.(\ref{st41}) is given  by $T$$=$$\textmd{diag}(t_{11},t_{22},t_{33})$ where $t_{11},t_{22},t_{33}$ are the eigen values of $\sqrt{\mathcal{W}^{T}\mathcal{W}},$ i.e., singular values of $\mathcal{W}.$
\subsection{$n$-local Linear Network}\label{km}
\begin{center}
\begin{figure}
\includegraphics[width=3.8in]{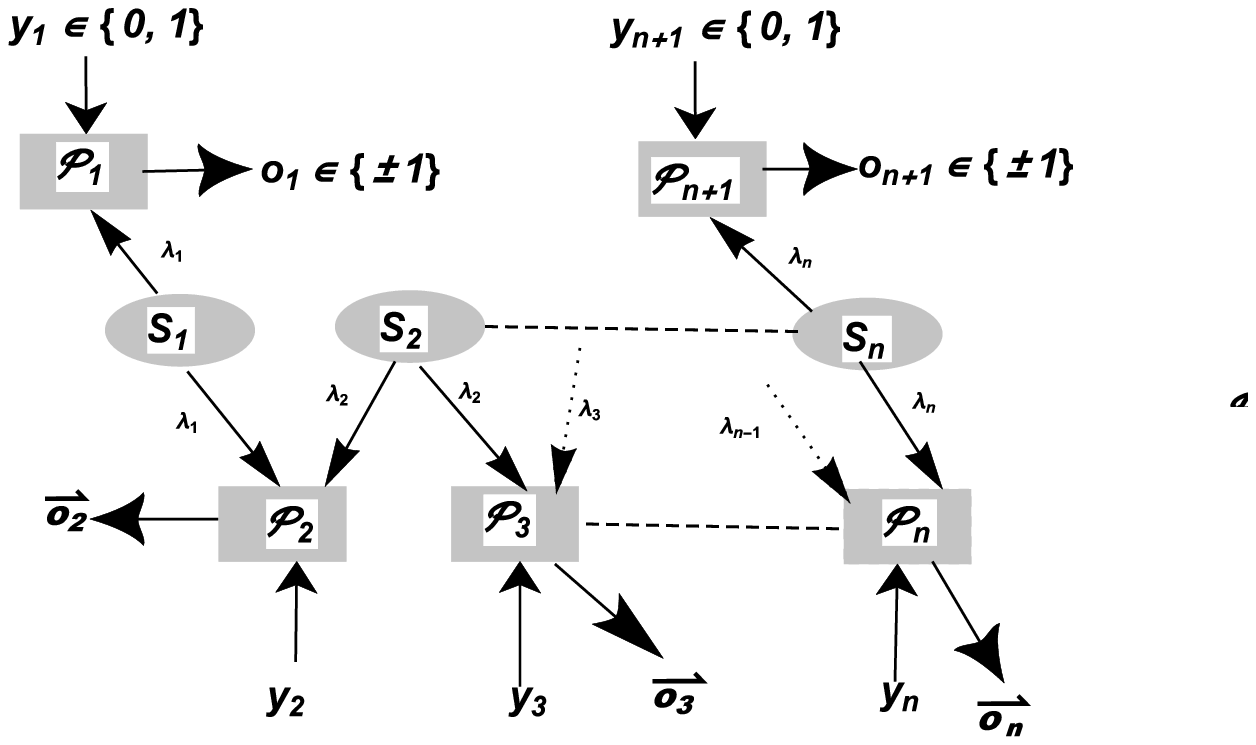} \\
 \caption{\emph{ Schematic representation of linear $n$-local network}}
\label{fig}
 \end{figure}
 \end{center}

Consider $n$ independent sources $\mathcal{S}_1,\mathcal{S}_2,...\mathcal{S}_n$ together with $n+1$ distant parties $\mathcal{P}_1,\mathcal{P}_2,...,\mathcal{P}_{n+1}$ arranged in a linear fashion (see Fig.\ref{fig}).
$\forall i$$=$$1,2,...,n,$ $\mathcal{S}_i$ independently distributes physical systems to a pair of parties $(\mathcal{P}_i,\mathcal{P}_{i+1}).$  Each of $\mathcal{P}_1$ and $\mathcal{P}_{n+1}$ receives particle from a single source $\mathcal{S}_1$ and $\mathcal{S}_n$ respectively. So the network is an open type of network. A source $\mathcal{S}_i$ is characterized by variable $\lambda_i.$ The sources being independent, the joint distribution of $\lambda_1,...,\lambda_n$ is factorizable:
\begin{equation}\label{tr1}
    \rho(\lambda_1,...\lambda_n)=\Pi_{i=1}^n\rho_i(\lambda_i).
 \end{equation}
where $\forall i,\,\rho_i$ denotes the normalized distribution of $\lambda_i.$ Eq.(\ref{tr1}) represents the $n$-local constraint. For $n$$=$$2,$
Eq.(\ref{tr1})
reduces to the bilocal constraint.\\
$\forall i$$=$$2,3,...n-1$ party $\mathcal{P}_i$ performs fixed measurement $y_i$ on the joint state of two subsystems received from two sources $\mathcal{S}_{i-1}, \mathcal{S}_{i}.$
Each of $\mathcal{P}_1$ and $\mathcal{P}_{n+1}$ chooses from a collection of two dichotomous measurements. $n+1$
partite network correlations are local if those can be decomposed $\textmd{as}\,:$
\begin{eqnarray}\label{tr2}
&&\small{p(\mathfrak{o}_1,\vec{\mathfrak{o}}_2,...,\vec{\mathfrak{o}}_n,\mathfrak{o}_{n+1}|y_1,y_{n+1})}=\nonumber\\
&&\int_{\Lambda_1}\int_{\Lambda_2}...\int_{\Lambda_n}
 d\lambda_1d\lambda_2...d\lambda_n\,\rho(\lambda_1,\lambda_2,...\lambda_n) R\\
  \,\textmd{where}&&\nonumber\\
&&R=p(\mathfrak{o}_1|y_1,\lambda_1)\Pi_{i=2}^n p(\vec{\mathfrak{o}}_i|\lambda_{i-1},\lambda_i) p(\mathfrak{o}_{n+1}|y_{n+1},\lambda_n)\nonumber
\end{eqnarray}
Notations appearing in the above expression are detailed below:
\begin{itemize}
  \item $\forall i,$ $\Lambda_i$ denotes the set of all possible values of local hidden variable $\lambda_i.$
  \item $y_1,y_{n+1}$$\in\{0,1\}$ denote measurements of $\mathcal{P}_1$ and $\mathcal{P}_{n+1}$ respectively.
  \item $\mathfrak{o}_1,\mathfrak{o}_{n+1}$$\in$$\{0,1\}$ denote outputs of $\mathcal{P}_1$ and $\mathcal{P}_{n+1}$ respectively.
  \item $\forall i,$ $\vec{\mathfrak{o}}_i$$=$$(\mathfrak{o}_{i1},\mathfrak{o}_{i2})$ denotes four outputs of input $y_i$ for $\mathfrak{o}_{ij}$$\in$$\{0,1\}$

\end{itemize}
In this linear scenario, $n+1$ partite network correlations are $n$-local if those satisfy both Eqs.(\ref{tr1},\ref{tr2}). So any set of correlations that do not satisfy both
of these constraints are said to be non $n$-local. The $n$-local inequality\cite{km1} for this scenario is given by:
\begin{eqnarray}\label{ineq}
   && \sqrt{|I|}+\sqrt{|J|}\leq  1,\,  \textmd{where}\\
 && I=\frac{1}{4}\sum_{y_1,y_{n+1}}\langle O_1O_2^0.....O_n^0O_{n+1}\rangle\nonumber\\
 &&   J= \frac{1}{4}\sum_{y_1,y_{n+1}}(-1)^{y_1+y_{n+1}}\langle \small{O_1O_2^1...O_n^1O_{n+1}}\rangle\,\,\textmd{with} \nonumber\\
&&   \langle O_1O_2^i.....O_n^iO_{n+1}\rangle = \sum_{\mathcal{D}}(-1)^{\mathfrak{o}_1+\mathfrak{o}_{n+1}+\mathfrak{o}_{2i}+...\mathfrak{o}_{ni}}R_1,\,\,i=0,1\nonumber\\
&& \textmd{\small{where}}\,R_1=\small{p(\mathfrak{o}_1,\vec{\mathfrak{o}}_2,...,\vec{\mathfrak{o}}_n,\mathfrak{o}_{n+1}|y_1,y_{n+1})} \nonumber\\
&&\textmd{\small{and}}\, \mathcal{D}=\{\mathfrak{o}_1,\mathfrak{o}_{21},\mathfrak{o}_{22},...,\mathfrak{o}_{n1},\mathfrak{o}_{n2},\mathfrak{o}_{n+1}\}\nonumber
\end{eqnarray}
Violation of Eq.(\ref{ineq}) guarantees non $n$-local behavior of corresponding correlations.

\subsection{Triangle Network\cite{gis1}}\label{train}
\begin{center}
\begin{figure}
\includegraphics[width=3.8in]{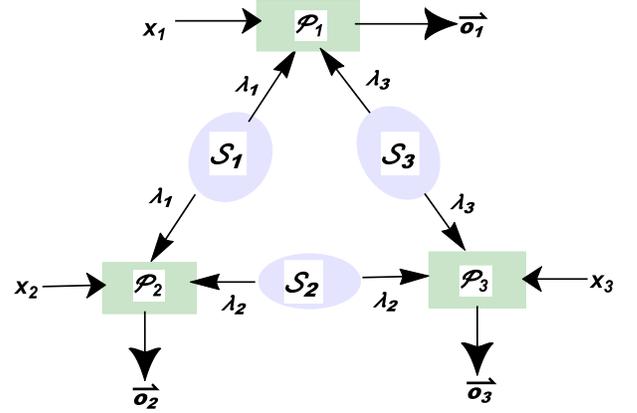} \\
 \caption{\emph{ Schematic representation of triangle network}}
\label{fig1}
 \end{figure}
 \end{center}

A triangle network is a non-linear arrangement of three parties $\mathcal{P}_1,\mathcal{P}_2,\mathcal{P}_3$ (say) and three independent sources
$\mathcal{S}_1,\mathcal{S}_2$ and $\mathcal{S}_3$ with the
parties corresponding to vertices of a triangle (see Fig.\ref{fig1}). Each source is placed between two parties. $\mathcal{S}_1, \mathcal{S}_2$
and $\mathcal{S}_3$ distribute particles to the pairs $(\mathcal{P}_1,\mathcal{P}_2),$ $(\mathcal{P}_2,\mathcal{P}_3)$ and $(\mathcal{P}_1,\mathcal{P}_3)$
respectively. Each party $\mathcal{P}_1,\mathcal{P}_2,\mathcal{P}_3$ thus receives two physical systems. Local variables $\lambda_i$ characterizes $\mathcal{S}_i$ for $i$$=$$1,2,3.$ The trilocal (source independence) assumption
takes the form:
\begin{equation}\label{tr3}
    \rho(\lambda_1,\lambda_2,\lambda_3)=\rho_1(\lambda_1)\rho_2(\lambda_2)\rho_3(\lambda_3)
\end{equation}
with $\rho_i$ corresponding to the normalized stochastic distribution of the local variable $\lambda_i, \forall i.$ Let $\textsl{x}_1,\textsl{x}_2$ and
$\textsl{x}_3$ denote the fixed local input of $\mathcal{P}_1,\mathcal{P}_2$ and $\mathcal{P}_3$ respectively. For $i$$=$$1,2,3,$ $\textsl{x}_i$ has four outputs denoted by a two component vector $\vec{\mathfrak{o}}_i$$=$$(\mathfrak{o}_{i1},\mathfrak{o}_{i2})$ where each component $\mathfrak{o}_{ij}$$\in$$\{0,1\}.$ Tripartite local correlations
 are of the form:
\begin{eqnarray}\label{tr4}
&&p(\vec{\mathfrak{o}}_1,\vec{\mathfrak{o}}_2,\vec{\mathfrak{o}}_3)=\int_{\Lambda_1}\int_{\Lambda_2}\int_{\Lambda_3} d\lambda_1d\lambda_2d\lambda_3 \rho(\lambda_1,\lambda_2,\lambda_3)R_2\nonumber\\
&&\\
&&\textmd{where}\,R_2=p(\vec{\mathfrak{o}}_1|\lambda_1,\lambda_3)p(\vec{\mathfrak{o}}_2|\lambda_1,\lambda_{2})p(\vec{\mathfrak{o}}_3|\lambda_2,\lambda_{3}).\nonumber
\end{eqnarray}
Any set of correlations generated in the network is trilocal if it is local, i.e., satisfies Eq.(\ref{tr4}) and also satisfies the trilocal
constraint given by Eq.(\ref{tr3}). Correlations lacking this type of representation (as specified by Eqs.(\ref{tr3},\ref{tr4})) are said to be \textit{nontrilocal}. \\
\section{Characterizing Trilocal Network Correlations}\label{main}
By construction, as specified in subsec.\ref{train}, the collection of trilocal correlations forms a proper subset of the network local correlations satisfying only Eq.(\ref{tr4}). Due to the nonlinear constraint (Eq.(\ref{tr3})), the set of trilocal correlations is non convex. A set of non-linear Bell-type inequalities is now framed. This set characterizes correlations in a triangle network in the sense that every member from the set is necessarily satisfied by trilocal correlations generated in the network.\\
\textit{Theorem.1:}\,In a triangle network, an arbitrary set of tripartite trilocal correlations necessarily satisfies:
\begin{equation}\label{tr5}
    \sqrt{|I_1|}+\sqrt{|I_2|}\leq 1,
\end{equation}
where correlator terms $I_1,I_2$ are specified as follows:
\begin{eqnarray}\label{tr6i}
  I_j &=& \sum_{i,k=0,1}(-1)^{j(i+k)}\langle C_1^i C_2^{j-1}C_3^k\rangle,\,\textmd{where}\nonumber\\
     \langle C_1^i C_2^{j-1}C_3^k\rangle &=& \sum_{\mathcal{C}}(-1)^{f(i,\vec{\mathfrak{o}_1})+
   g(j-1,\vec{\mathfrak{o}_2})+h(k,\vec{\mathfrak{o}_3})}p(\vec{\mathfrak{o}}_1,\vec{\mathfrak{o}}_2,\vec{\mathfrak{o}}_3)\nonumber\\
   &&
  \end{eqnarray}
with $j$$=$$1,2$, $\mathcal{C}$$=$$\{\mathfrak{o}_{11},\mathfrak{o}_{12},\mathfrak{o}_{21},\mathfrak{o}_{22},
   \mathfrak{o}_{31},\mathfrak{o}_{32}\}$ and $\mathfrak{o}_{ij}$$\in$$\{0,1\}.$\\
In the above expression, each of $f(.,.),g(.,.)$ and $h(.,.)$ denotes arbitrary integer-valued function.


\textit{Proof:}\,See Appendix.A\\
$\forall f,g,h,$ the criterion provided by Eq.(\ref{tr5}) is satisfied by trilocal correlations. So, Eq.(\ref{tr5}) gives a set of trilocal inequalities. Violation of at least one of these inequalities (for suitable form of $f,g,h$) ensures \textit{nontrilocality} of the measurement statistics. The corresponding triangle network is said to be \textit{nontrilocal}. Violation of at least one of the inequalities specified by Eq.(\ref{tr5}) thus acts as a detection criterion of nontrilocality in a triangle network. However, such a criterion being only sufficient for the purpose, there may exist nontrilocal correlations satisfying Eq.(\ref{tr5}) for all possible forms of $f,g,h$. Now, one may note that the labeling of the parties can be interchanged in Eq.(\ref{tr6i}) to get a different set of trilocal inequalities.\\
Having established a set of trilocal inequalities, quantum violation of the same is analyzed in the next section.\\
\section{Quantum Triangle Network}\label{quan}
Let each source $\mathcal{S}_i$ distribute a two qubit state $\varrho_i\,(i$$=$$1,2,3)$ Each of $\mathcal{P}_1,\mathcal{P}_2$ and
$\mathcal{P}_3$ receives two qubits (see Fig.\ref{fig1}). To be specific, $ \varrho_1$ gets distributed among parties $\mathcal{P}_1,\mathcal{P}_2;$ $ \varrho_2$ gets distributed among $\mathcal{P}_2,\mathcal{P}_3$ and $ \varrho_3$ gets distributed among $\mathcal{P}_1,\mathcal{P}_3;$ Overall quantum state in the network is thus given by:
\begin{equation}\label{tr7}
    \varrho_{123}=\varrho_1\otimes\varrho_2\otimes\varrho_3
\end{equation}
Each of the parties now performs a fixed local quantum measurement on the joint state of their respective subsystems (two qubits). In case the measurement statistics
violate the trilocal inequality (Eq.(\ref{tr5})), nontrilocality is observed in the quantum network. However, no such definite conclusion can be made
otherwise. In a triangle network, a six qubit state $\varrho_{123}$ is distributed among the three parties where each party is receiving two qubits from two different bipartite states. In case violation is observed, it may be attributed as \textit{nontrilocality} of $\varrho_{123}$ (Eq.(\ref{tr7})).
\subsection{Quantum Violation}\label{viol}
To illustrate quantum violation, one of the instances given in \cite{gis1} is first considered. Let $\mathcal{S}_1,\mathcal{S}_2,\mathcal{S}_3$ each
generate a Bell state:
\begin{equation}\label{tr8}
    \varrho_i=\frac{1}{\sqrt{2}}(|00\rangle+|11\rangle),\,\,i=1,2,3
\end{equation}
 On receiving the qubits, each of $\mathcal{P}_1,\mathcal{P}_2$ and $\mathcal{P}_3$ performs measurement on the following orthonormal basis:
\begin{eqnarray}\label{tr9}
  |b_1\rangle=|01\rangle,\,\,  |b_2\rangle=|10\rangle,\,\,|b_3\rangle=\alpha_1|00\rangle+\alpha_2|11\rangle\, \textmd{and}\quad\quad&&\nonumber \\
 |b_4\rangle=\alpha_2|00\rangle-\alpha_1|11\rangle,\,\textmd{\small{with}}\,\small{0<\alpha_2,\alpha_1<1}\,\textmd{\small{and}}\,\alpha_1^2+\alpha_2^2=1&&\nonumber\\
\end{eqnarray}

Now violation of Eq.(\ref{tr5}) for any specific form of the integer-valued functions $f,g,h$ ensures nontrilocality of corresponding correlations. Let us consider the following integer-valued functions:
\begin{eqnarray}\label{tr6iv}
    f(s,\vec{r}) &=& g(s,\vec{r})=(1-s)r_1+s(r_2+1)\, \textmd{and}\nonumber\\
  h(s,\vec{r}) &=& (1-s)|r_1+r_2-1|+s|r_1r_2-1|\\
  \textmd{where}\,\,\vec{r}&=&(r_1,r_2)\,\textmd{and}\,\,s,r_1,r_2\in\{0,1\} \nonumber
\end{eqnarray}
In our discussions, we will refer to the specific trilocal inequality obtained using the above functions by Eqs.(\ref{tr5},\ref{tr6iv}).

\begin{center}
\begin{figure}
\includegraphics[width=2.7in]{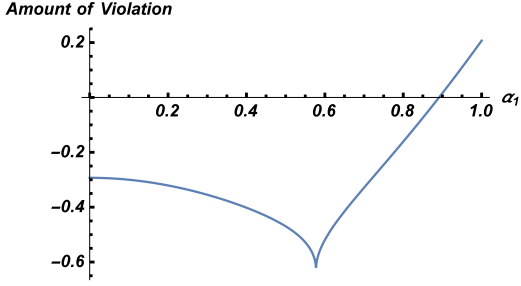} \\
 \caption{\emph{ Plotting range of measurement parameter $\alpha_1$ for which Eq.(\ref{tr10k}) is violated assuming $\alpha_2$$=$$\sqrt{1-\alpha_1^2}$ (Eq.(\ref{tr9})). In this case three identical copies of a pure entangled state (Eq.(\ref{tr8})) are used in the network.}}
\label{fig3}
 \end{figure}
 \end{center}

For corresponding measurement statistics, Eqs.(\ref{tr5},\ref{tr6iv}) give:
\begin{equation}\label{tr10k}
\sqrt{|2\alpha_1^2-\alpha_2^2|}+\sqrt{|\alpha_1^6 - \alpha_1^4 \alpha_2^2 +\alpha_2^6 + \alpha_1^2 (3 - \alpha_2^4)|}\leq 2^{\frac{3}{2}}
\end{equation}
Relation (Eq.(\ref{tr10k})) between the measurement parameters $\alpha_1,\alpha_2$ points out that violation of Eqs.(\ref{tr5},\ref{tr6iv}) is obtained for
$\alpha_1$$>$$0.892$ (see Fig.\ref{fig3}).
Consequently, nontrilocality is observed in the network.
\textit{Reproducing Result of \cite{gis1}:} In \cite{gis1}, the authors provided an instance of nontrilocal correlations. They used
a Bell state (Eq.(\ref{tr8})) and the measurement contexts given by Eq.(\ref{tr9}). Precisely speaking, they claimed
that there does not exist any trilocal model explaining the correlations arising for $0.785$$<$$\alpha_1^2$$<$$1,$ i.e., for $\alpha_1$$\in$$(0.886,1).$ Now, the correlations violate Eqs.(\ref{tr5},\ref{tr6iv}) for $\alpha_1$$\in$$(0.892,1).$ So excepting the small gap $(0.886,0.892],$
the trilocal inequality (Eqs.(\ref{tr5},\ref{tr6iv})) successfully detects nontrilocality as was ensured in \cite{gis1} by proving non existence of any trilocal model.\\
\subsection{Nontrilocal Correlations In Network Involving Separable States And/Or Measurement In Product State Basis}
Let us first consider a triangle network where each of the three sources generate an identical copy of a pure entangled state (Eq.(\ref{tr8})). Each party ($\mathcal{P}_i$)
performs joint measurement (on the state of their respective two qubits received) in the following product state basis:
\begin{eqnarray}\label{prodbas}
  |b_1\rangle=|01\rangle, && |b_2\rangle=|10\rangle,\nonumber \\
  |b_3\rangle=|11\rangle, && |b_4\rangle=|00\rangle.
\end{eqnarray}
For instance, consider a trilocal inequality (Eq.(\ref{tr5})) specified by the following functions:
\begin{eqnarray}\label{tr6}
    f(s,\vec{r}) &=& (1-s)|r_1+r_2-1|+s|r_1r_2-1|\, \textmd{and}\nonumber\\
 g(s,\vec{r})&=& h(s,\vec{r}) =(1-s)r_1+s(r_2+1) \\
 \textmd{where}\,\, \vec{r}&=&(r_1,r_2)\,\textmd{with}\,\,r_1,r_2,s\in\{0,1\}\nonumber
\end{eqnarray}

Correlator terms in the inequality (Eqs.(\ref{tr5},\ref{tr6})) become:
\begin{equation}
I_1= \frac{1}{4}\,\textmd{and}\, I_2=\frac{1}{2}.
\end{equation}
Clearly, violation of the inequality (Eqs.(\ref{tr5},\ref{tr6})) is observed$.\,$Tripartite correlations generated in the network thus turn out to be nontrilocal. \\
Next, let each source generate an identical copy of a separable state:
\begin{equation}\label{prodbas1}
\varrho_{sep}=\frac{1}{2}(|00\rangle\langle 00|+|11\rangle\langle 11|)
\end{equation}
Let each party now measure on the orthonormal basis given by:
\begin{eqnarray}\label{tr9new}
|b_1\rangle= \alpha_1|01\rangle+\alpha_2|10\rangle, && |b_2\rangle=\alpha_2|01\rangle-\alpha_1|10\rangle,\nonumber\\
|b_3\rangle=\alpha_3|00\rangle+\alpha_4|11\rangle, && |b_4\rangle=\alpha_3|00\rangle-\alpha_4|11\rangle
\end{eqnarray}
with  $0\leq\alpha_1,\alpha_2,\alpha_3,\alpha_4$$\leq$$1$ and $\alpha_1^2+\alpha_2^2$$=$$\alpha_3^2+\alpha_4^2$$=$$1.$

\begin{center}
\begin{figure}
\includegraphics[width=2.7in]{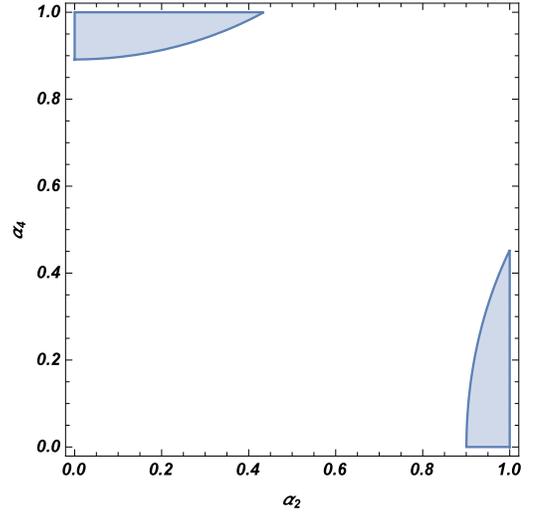} \\
 \caption{\emph{ Shaded region indicates the measurement parameters $\alpha_2,\alpha_4$ of the orthogonal basis (Eq.(\ref{tr9new})) for which nontrilocal correlations are observed in triangle network involving identical copies of a separable state (Eq.(\ref{prodbas1})).}}
\label{fig5}
 \end{figure}
 \end{center}

 Violation of the trilocal inequality (Eqs.(\ref{tr5},\ref{tr6})) is observed if:
\begin{eqnarray}\label{prodbas2}
\sqrt{|1 + \alpha_2^2 + 2 \alpha_2^4 - \alpha_4^2 - 6 \alpha_2^2 \alpha_4^2 + 4 \alpha_4^4|}+\nonumber\\
\sqrt{|(1-2 \alpha_2^2) (1 + \alpha_2^2 - 3 \alpha_4^2)|}>2^{\frac{3}{2}}.
\end{eqnarray}
There exists a range of the measurement parameters $\alpha_2,\alpha_4$ for which the above relation is true (see Fig.\ref{fig5}). So, nontrilocality is obtained even if the network involves separable states.\\
Now for $(\alpha_2,\alpha_4)$$=$$(1,0)$ or $(0,1),$ the basis given by Eq.(\ref{tr9new}) is a product state basis. Also, one may note that violation is observed for these particular values of the measurement parameters. This in turn justifies that nontrilocal correlations are generated in a triangle network involving separable states and/or measurement on a product state basis. In this context, it may be noted that the same is not true in case each source generates a product state.
Precisely speaking, for no form of integer-valued functions $f,g,h,$ the trilocal inequality (Eq.(\ref{tr5})) is violated in a triangle network involving two-qubit product states. The illustration follows below.\\
\subsection{No Detection of Non trilocality in Network Involving Product States}\label{product}
$\forall i,$ let $\mathcal{S}_i$ distribute a two-qubit product state:
\begin{eqnarray}\label{prodbas4}
    \varrho_{prod}^{(i,i+1)}&=&\frac{1}{4}(\mathbb{I}_{2}+\vec{u}_i.\vec{\sigma})(\mathbb{I}_{2}+\vec{u}_{i+1}.\vec{\sigma})\\
    \textmd{where}&&i=1,3,5\,\textmd{and}\,||\vec{u}_i||,||\vec{u}_{i+1}||\leq 1.\nonumber
\end{eqnarray}
The overall state in the network is thus given by:
\begin{equation}\label{prodbas5}
\varrho_{prod}=\frac{1}{2^6}\Pi_{i=1}^6(\mathbb{I}_{2}+\vec{u}_i.\vec{\sigma}).
\end{equation}
As per the arrangement of the parties discussed in the subset.\ref{train}, party $\mathcal{P}_1$ receives two qubits forming the product state:
$\frac{1}{4}(\mathbb{I}_{2}+\vec{u}_1.\vec{\sigma})(\mathbb{I}_{2}+\vec{u}_{6}.\vec{\sigma}).$ Similarly $\mathcal{P}_2$ and $\mathcal{P}_3$ have:
$\frac{1}{4}(\mathbb{I}_{2}+\vec{u}_2.\vec{\sigma})(\mathbb{I}_{2}+\vec{u}_{3}.\vec{\sigma})$ and $\frac{1}{4}(\mathbb{I}_{2}+\vec{u}_4.\vec{\sigma})(\mathbb{I}_{2}+\vec{u}_{5}.\vec{\sigma})$ respectively.
Let $M_i$ denote the single measurement (with four possible outputs) performed by $\mathcal{P}_i\,(i$$=$$1,2,3)$ on its respective two-qubit state. Let
$\vec{\mathfrak{o}_i}$$=(\mathfrak{\mathfrak{o}_{i1}},\mathfrak{\mathfrak{o}_{i2}})$ with $\mathfrak{\mathfrak{o}_{ij}}$$\in$$\{0,1\}$  denote the four outputs of $M_i.$
Clearly, in this case, each probability term $p(\vec{\mathfrak{o}_1},\vec{\mathfrak{o}_2},\vec{\mathfrak{o}_3})$ is factorizable. So,
\begin{equation}\label{prodbas6}
     \langle C_1^i C_2^{j-1}C_3^k\rangle = \langle C_1^i\rangle \langle C_2^{j-1}\rangle \langle C_3^k\rangle ,\,\,\forall i,j,k.
\end{equation}
Using the above relation, we get:
\begin{eqnarray}\label{prodbas8}
  I_1&=&\frac{\langle C_1^0-C_1^1\rangle\langle C_2^0\rangle \langle C_3^0-C_3^1\rangle}{4} \nonumber\\
  I_2&=&\frac{\langle C_1^0+C_1^1\rangle\langle C_2^1\rangle \langle C_3^0+C_3^1\rangle}{4}.
   \end{eqnarray}
For $M_2,$ the corresponding correlator term $\langle C_2^j\rangle$ is given by:
\begin{equation}\label{aver1}
\langle C_2^j\rangle= \sum_{\mathfrak{o}_{21},\mathfrak{o}_{22}}(-1)^{g(j,\vec{\mathfrak{o}_2})}p(\vec{\mathfrak{o}}_2)   ,
\end{equation}
where $j$$=$$0,1$ and $g(j,\vec{\mathfrak{o}_2})$ is any integer-valued function. Eq.(\ref{aver1}) implies that $|\langle C_2^j\rangle|$$\leq$$1.$ So   Eq.(\ref{prodbas8}) gives:
\begin{equation}\label{prodbas9i}
    \sqrt{|I_1|}+\sqrt{|I_2|}\leq \frac{1}{2}\sum_{t=0}^1 \sqrt{\Pi_{j=1,3}|\sum_{i=0}^1 \langle (-1)^{it} C_j^i\rangle|}.
\end{equation}
Now for any $4$ positive numbers, $k_1,k_2,k_3,k_4$, we have:
\begin{equation}\label{prodbas7}
    \sqrt{k_1k_2}+\sqrt{k_3k_4}\leq\sqrt{k_1+k_3}\sqrt{k_2+k_4}.
\end{equation}
Using Eq(\ref{prodbas7}), we get from Eq.(\ref{prodbas9i}):
\begin{eqnarray}\label{prodbas9}
    \sqrt{|I_1|}+\sqrt{|I_2|}&\leq& \Pi_{j=1,3}\sqrt{\frac{\sum_{i=0}^1| \langle C_j^0\rangle +(-1)^{i} \langle C_j^1\rangle|}{2}}\nonumber\\
    &=&\Pi_{j=1,3}\sqrt{\small{\textmd{max}}\{|\langle C_j^0\rangle|,|\langle C^1_j\rangle|\}}\nonumber\\
    &\leq& 1
\end{eqnarray}
Hence, none of the trilocal inequalities from the set (Eq.(\ref{tr5})) shows a violation. Consequently, non trilocality is not detected by Eq.(\ref{tr5}) when each source distributes a product state in the network. Such an observation points out that starting with no correlation between any pair of parties, non trilocality cannot be detected if the parties are not allowed to interact among themselves in the network.
\par Based on the above analysis of quantum violation of the trilocal inequality (Eq.(\ref{tr5})), a pure bipartite entanglement detection scheme is designed next.
\section{Detection of Pure Entanglement}\label{ent}
Consider the particular trilocal inequality given by Eqs.(\ref{tr5},\ref{tr6}) in a triangle network. Let each of the three sources $\mathcal{S}_i$ generate an unknown pure bipartite state $\kappa_i$ ($i$$=$$1,2,3).$  Qubits of the states are distributed to the parties. Each of $\mathcal{P}_1,\mathcal{P}_2$ and $\mathcal{P}_3$ performs a measurement in the basis given by Eq.(\ref{tr9}). The corresponding tripartite correlations are used to test the trilocal inequality (Eqs.(\ref{tr5},\ref{tr6})). Violation of this inequality ensures that each of $\kappa_1,\kappa_2$ and $\kappa_3$ is entangled. Such a deduction is based on the observation that violation of this inequality is impossible if at least one of the three sources generates a pure product state. The justification of the claim is stated below.\\
It is already discussed in subsec.\ref{product} that if all the states used in the network are product states, then there is no violation of any trilocal inequality from the set given by Eq.(\ref{tr5}). So, given the information that all the sources generate a pure state, for violation of Eqs.(\ref{tr5},\ref{tr6}), at least one of those pure states must be entangled. Let only one source generate a product state (Eq.(\ref{prodbas4}) for $||\vec{u}_i||,||\vec{u}_{i+1}||$$=$$1$) and each of the remaining sources generate a pure entangled state\cite{nie}:
\begin{equation}\label{ent1}
    \varrho_{ent}^{i}=\frac{1}{\sqrt{2}}(\tau_{1i}|00\rangle+\tau_{2i}|11\rangle)\,\,\textmd{and}\, \tau_{1i}^{2}+\tau_{2i}^2=1.
\end{equation}
In Eq.(\ref{ent1}), $i$ denotes labeling of the source generating the corresponding state ($\mathcal{S}_i$ generating $\varrho_{ent}^{i}$).
The constraint required for a violation of the trilocal inequality (Eqs.(\ref{tr5},\ref{tr6})) for all possible cases (depending on which of the two sources distribute pure entanglement) are given in Table.\ref{table:ta1} in Appendix.B. By numerical maximization it is observed that the upper bound of Eqs.(\ref{tr5},\ref{tr6}) does not exceed $1$ for all possible values of the state and measurement parameters. So, violation of the particular inequality (Eqs.(\ref{tr5},\ref{tr6})) is impossible even if two of the three states involved in the network are entangled. However, it is evident from discussions in subsec.\ref{viol} that the same inequality can be violated when all the sources generate pure entanglement. Consequently, in case of violation is observed in the network involving only pure states, all of them must be entangled. However, in case the sources distribute mixed states, no definite conclusion can be drawn on observing the violation of the same inequality.\\

\section{Resistance To Noise}\label{noise}
A detection criterion provided by any of the trilocal inequality (Eq.(\ref{tr5})) is based on correlators. From experimental perspectives, it thus becomes important to discuss noise tolerance of any such criterion. In practical situations, testing any such inequality (Eq.(\ref{tr5})) may encounter different types of noise. Here illustration is provided for some of the possible cases and considering some particular trilocal inequalities, i.e., considering Eq.(\ref{tr5}) for some specific forms of $f,g,h$.
\subsection{Error in Entanglement Generation}
\begin{center}
\begin{figure}
\includegraphics[width=2.7in]{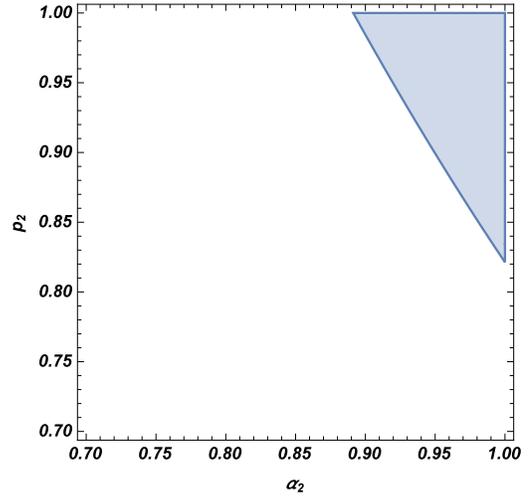} \\
 \caption{\emph{ Subspace of the parameter space $(\alpha_2,p_2)$ characterizing error (in entanglement generation) tolerance of the trilocal inequality
 (Eqs.(\ref{tr5},\ref{tr6})) is shown here.}}
\label{fig6}
 \end{figure}
 \end{center}

Let us first consider the ideal situation$.\,$Initially, each of $\mathcal{S}_1,\mathcal{S}_2$ and $\mathcal{S}_3$ has the state $\varrho$$=$$|01\rangle\langle 01|.$  Application of the Hadamard gate ($\mathcal{H}$) on the first qubit followed by the C$\small{-}$NOT ($\mathcal{C}\small{-}\mathcal{NOT}$) gate (with the first qubit as control qubit) results in the generation of the Bell state $|\phi^-\rangle\langle \phi^-|$\cite{indr1}. In practical situations, while generating the entangled state, imperfections may occur in applications of both one and two-qubit operations. At each source, let $p_1$ and $p_2$ denote the imperfection parameters characterizing $\mathcal{H}$ and $\mathcal{C}\small{-}\mathcal{NOT}$ respectively. After application of a noisy Hadamard gate, state ($\varrho$) becomes\cite{indr5}:
\begin{eqnarray}\label{had1}
\varrho^{'}&=&p_1( \mathcal{H}\otimes\mathbb{I}_2 \varrho \mathcal{H}^{\dagger}\otimes\mathbb{I}_2)+\frac{1-p_1}{2}\mathbb{I}_2\otimes\varrho_2,\nonumber\\
&&\textmd{where}\,\, \varrho_2=\textmd{Tr}_1(\varrho).\nonumber\\
&=&\frac{1}{2}(|00\rangle\langle 00|+|10\rangle\langle 10|)-\frac{p_1}{2}(|00\rangle\langle 10|+|10\rangle\langle 00|)\nonumber\\
&&
\end{eqnarray}
When subjected to noisy $\mathcal{C}\small{-}\mathcal{NOT},$ $\varrho^{'}$ becomes\cite{indr5}:
\begin{eqnarray}\label{had2}
 \varrho^{''}&=&p_2( \mathcal{C}\small{-}\mathcal{NOT}\varrho^{'}(\mathcal{C}\small{-}\mathcal{NOT})^{\dagger})+\frac{1-p_2}{4}\mathbb{I}_2\otimes \mathbb{I}_2\nonumber\\
&=&\frac{1}{4}(\sum_{i,j=0}^1(1+(-1)^{i+j}p_2)|ij\rangle\langle ij|-2p_1p_2(|11\rangle\langle 00|+\nonumber\\
&&|00\rangle\langle 11|))
\end{eqnarray}
In realistic situations, each source distributes $\varrho^{''}$(Eq.(\ref{had2})) in the network$.\,$On receiving the qubits, each of the parties measures in the basis given by Eq.(\ref{tr9}). Now, let us consider the particular trilocal inequality given by Eqs.(\ref{tr5},\ref{tr6}). In an ideal scenario, this particular trilocal inequality detects nontrilocal correlations in the network. In a triangle network involving noisy states (Eq.(\ref{had2})), the measurement statistics violate the inequality if:
\begin{eqnarray}\label{had3}
\sqrt{|p_2(p_2 - 3 p_2 \alpha_2^2 - (1 - p_2) \alpha_2^4)|} +&&\nonumber\\
\sqrt{|p_2^2 (p_2 - \alpha_2^2 + 4 \alpha_2^4)|} >2^{\frac{3}{2}}.&&
\end{eqnarray}
Above relation (Eq.(\ref{had3})) is satisfied for some range of $p_2$ and the measurement parameter $\alpha_2$ (see Fig.\ref{fig6}). This in turn illustrates the noise tolerance of the trilocal inequality (Eqs.(\ref{tr5},\ref{tr6})).
\subsection{Communication Over Noisy Quantum Channels}
\begin{center}
\begin{figure}
\includegraphics[width=2.7in]{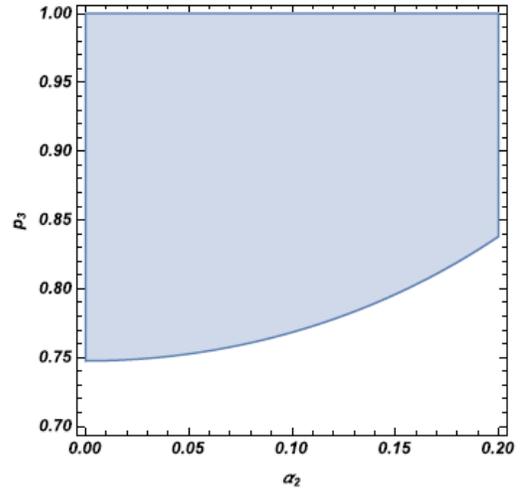} \\
 \caption{\emph{ Shaded region indicates the measurement parameter ($\alpha_2$) and depolarizing noise parameter ($p_3$) for which nontrilocal correlations are observed in triangle network involving identical copies of the noisy state $\varrho_{dep}$ (Eq.(\ref{had4})).}}
\label{fig7}
 \end{figure}
 \end{center}

 Now let each source generate a pure entangled state (Eq.(\ref{tr8})). However, the states received by the parties are noisy due to the communication of one or both qubits (from sources in the network) over noisy quantum channels. For instance, let the bipartite state (from each source) be passed through a  depolarizing channel characterized by noise parameter $p_3$ (say). Let depolarization occurs with probability $1$$-$$p_3$\cite{nie}. Identical copies of the noisy class of states involved in the network are given by:
\begin{equation}\label{had4}
\varrho_{dep}=p_3 |\phi^-\rangle\langle \phi^-|+\frac{1-p_3}{4}\mathbb{I}_4.
\end{equation}
Let each of the parties perform a measurement in the measurement basis given by Eq.(\ref{tr9}). The trilocal inequality (Eqs.(\ref{tr5},\ref{tr6})) gives:
\begin{equation}\label{toler}
    |p_3^2 (\alpha_2^2 - 4 \alpha_2^4 + p_3)|+|p_3 (p_3 - 3 \alpha_2^2 p_3 + \alpha_2^4 (1 + p_3))|\leq 2^{\frac{3}{2}}.
\end{equation}
Numerical maximization of the left-hand side expression of the above inequality yields $0.568$ approximately. Hence, Eq.(\ref{toler}) is satisfied for all possible values of $p_3,\alpha_2.$ This in turn implies that the particular trilocal inequality (Eqs.(\ref{tr5},\ref{tr6})) has zero noise tolerance in this case. So no violation is observed$.\,$Non trilocality can be detected over some range of $p_3,$ considering some other trilocal inequality from the set given by Eq.(\ref{tr5}). For example, let us consider the following functions in Eq.(\ref{tr5}):
\begin{eqnarray}\label{had5}
 f(s,\vec{r})&=&1\,\textmd{if}\,s=1\,\textmd{and}\,r_1r_2=0\nonumber \\
 &=&0\,\textmd{else}\nonumber\\
 g(s,\vec{r})&=& h(s,\vec{r})=(1-s)|r_1+r_2-1|+s|r_1r_2-1|\, \textmd{with}\nonumber\\
  \vec{r}&=&(r_1,r_2)\,\textmd{and}\,s,r_1,r_2\in\{0,1\}
\end{eqnarray}
Clearly all these functions(Eq.(\ref{had5})) are integer-valued. Eq.(\ref{tr5}) together with these functional forms, therefore, gives a trilocal inequality. So all trilocal correlations necessarily satisfies the inequality  given by Eqs.(\ref{tr5},\ref{had5}). Its violation thus guarantees the generation of nontrilocal correlations in the triangle network. Now
 violation of this particular trilocal inequality is observed if:
\begin{eqnarray}\label{had6}
    \sqrt{|W_1|}+2\sqrt{|W_2|}>8\qquad\qquad\quad\qquad\quad&&\\
 \textmd{with}\,\, \small{W_1}=2 + 6 p_3 + 2 p_3^2 - 2 p_3^3 + 8 \alpha_2^4 p_3 (1 + p_3)\qquad\qquad&& \nonumber\\
  - 8 \alpha_2^2 p_3 (2 + p_3)\,\,\textmd{and}\qquad\qquad\qquad\qquad\qquad&&\nonumber\\
    W_2=p_3 (9 + \alpha_2^2 (-22 - 4 p_3) + p_3 + \alpha_2^4 (14 + 6 p_3)).\qquad\quad&&\nonumber
\end{eqnarray}
Above relation is satisfied  for some range of $p_3$ (see Fig.\ref{fig7}).\\
\subsection{Imperfect Measurements}
\begin{center}
\begin{figure}
\includegraphics[width=2.7in]{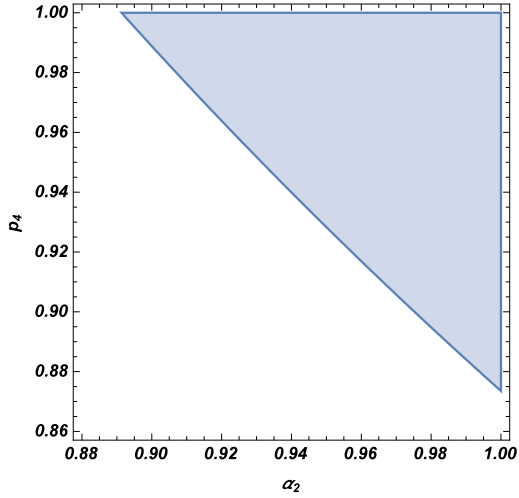} \\
 \caption{\emph{ Shaded region specifies the ideal measurement parameter ($\alpha_2$) and imperfection parameter ($p_4$) for which nontrilocal correlations are detected in a triangle network involving identical copies of Bell state $|\phi^+\rangle\langle \phi^+|.$}}
\label{fig8i}
 \end{figure}
 \end{center}

Detection inefficiency is another potent factor of the experimental noise which forms the basis of detection loophole\cite{b33}. Let each of the parties perform a measurement in the orthogonal basis given by Eq.(\ref{tr9}) with detection efficiency $p_4.$ To be precise, on being measured in the basis (Eq.(\ref{tr9})), the output is detected efficiently with probability $p_4$ whereas no output is obtained with probability $1$$-$$p_4.$ The corresponding POVM elements of such imperfect basis measurement are given by $p_4 b_i+\frac{1-p_4}{4}\mathbb{I}_4$ where $b_1,b_2,b_3,b_4$ are the elements of basis given by Eq.(\ref{tr9}). In case identical copies of the pure entangled state($|\phi^+\rangle\langle \phi^+|$) are used in the network, violation of the trilocal inequality given by Eqs.(\ref{tr5},\ref{tr6}) is observed if:
\begin{eqnarray}\label{had7}
 \sqrt{|p_4^2 (3 \alpha_2^2 p_4-\alpha_2^4 (1 - p_4) - p_4)|}+&&\nonumber\\
  \sqrt{|(1 -\alpha_2^2 +4 \alpha_2^4) p_4^3|} >2^{\frac{3}{2}}.&&
\end{eqnarray}
Over some restricted range of $p_4$ (see Fig.\ref{fig8i}), nontrilocal correlations are detected via violation of the trilocal inequality (Eqs.(\ref{tr5},\ref{tr6})).\\

\section{Comparison With Existing Trilocal Inequality (Eq.(\ref{ineq}))}\label{comp}
There exist dissimilarities between a linear trilocal network (Fig.\ref{fig}) and a triangle network (Fig\ref{fig1})$.\,\,$Structural dissimilarity between an open type (linear trilocal network) and a closed form of network (triangle network) is obvious. Apart from that, it may be noted that while the measurement scenario involved in a triangle network is a fixed input scenario, not all parties in a linear network perform fixed measurement (subsec.\ref{km}). As already discussed before, under fixed input assumption, nonclassical network correlations can be simulated which are not attributable to the standard Bell nonlocality.  To that end, it becomes imperative to compare the trilocal inequalities for a triangle network with the existing trilocal inequality (Eq.(\ref{ineq})) for a linear network\cite{km1}.
\subsection{Detecting Nontrilocality in Triangle Network Involving Bell-CHSH local States}
Let us first consider the trilocal linear network. Here each of the three independent sources distributes two-qubit states among four parties $\mathcal{P}_1,\mathcal{P}_2,\mathcal{P}_3$ and $\mathcal{P}_4.$
Each of the two parties $\mathcal{P}_2,\mathcal{P}_3,$ receiving two qubits, performs Bell basis measurement. Each of the remaining two parties $\mathcal{P}_1,\mathcal{P}_4,$ receiving single qubit, performs local projective measurements.
In case an arbitrary two qubit state $\varrho_i$ (Eq.(\ref{st4})) is generated by source $\mathcal{S}_i\,(i$$=$$1,2,3)$, nontrilocal correlations are detected by violation of Eq.(\ref{ineq}) if\cite{bilo5}:
\begin{equation}\label{tribd}
    \sqrt{\Pi_{i=1}^3t_{i11}+\Pi_{i=1}^3t_{i22}}>1
\end{equation}
where $t_{i11},t_{i22}$ denote the largest two singular values of correlation tensor ($T_i$,say) of $\varrho_i\,(i$$=$$1,2,3).$ So in case Eq.(\ref{tribd})
does not hold, the correlations turn out to be trilocal as per the trilocal inequality (Eq.(\ref{ineq})). Now, let us consider the case where all the three states ($\varrho_1,\varrho_2,\varrho_3$) are Bell-CHSH local\cite{hor}:
\begin{equation}\label{horcr}
    \sqrt{t_{i11}^2+t_{i22}^2}\leq 1,\,\, \forall i=1,2,3.
\end{equation}
Again,
\begin{eqnarray}\label{chsh1}
  \Pi_{i=1}^3t_{i11}+\Pi_{i=1}^3t_{i22} \leq \sqrt{AB}\qquad&&\\
  \textmd{where,}\, A=\Pi_{i=1}^2t_{i11}^2+\Pi_{i=1}^2t_{i22}^2&&\nonumber\\
   \textmd{and}\, B=t_{311}^2+t_{322}^2.\qquad\qquad\quad&&\nonumber
\end{eqnarray}
Using Eq.(\ref{horcr}) for $i$$=$$1,2,$ we get:
\begin{eqnarray}\label{chsh2}
   \sqrt{\Pi_{i=1}^2t_{i11}^2+\Pi_{i=1}^2t_{i22}^2+t_{111}^2t_{222}^2+t_{122}^2t_{211}^2} && \nonumber\\
   = \sqrt{\Pi_{i=1}^2t_{i11}^2+\Pi_{i=1}^2t_{i22}^2+C}\,\,\,,\,\,\qquad\qquad&&\nonumber\\
   \leq1\,\,\,\,\,\,\,\,\,\,\,\,\,\,\,\quad\qquad\qquad\qquad\qquad\qquad\qquad&&\\
   \textmd{where,}\,\, C=t_{111}^2t_{222}^2+t_{122}^2t_{211}^2.\qquad\qquad&&\nonumber
\end{eqnarray}
Using Eq.(\ref{chsh2}), expression $\sqrt{\Pi_{i=1}^2t_{i11}^2+\Pi_{i=1}^2t_{i22}^2}$ in Eq.(\ref{chsh1}) turns out to be $\leq$$1.$ Also as Eq.(\ref{horcr})
holds for $i$$=$$3,$ Eq.(\ref{chsh1}) gives:
\begin{equation}\label{chsh3}
     \Pi_{i=1}^3t_{i11}+\Pi_{i=1}^3t_{i22} \leq 1.
\end{equation}
So, in case each state used in the network is Bell-CHSH local, corresponding network correlations are not detected as non trilocal by Eq.(\ref{ineq}). Consequently, the violation of Eq.(\ref{ineq}) relies on Bell-CHSH inequality violation by the
states distributed in the network. However, in the triangle network, in some cases, it is observed that for a suitable choice of
the integer-valued functions ($f,g,h$), violation of a trilocal inequality (Eq.(\ref{tr5})) is
possible even if all the sources distribute Bell-CHSH local states.\\
Consider the family of Bell diagonal states\cite{nie}:
\begin{eqnarray}\label{belldiag}
    \varrho_{Bell}=\omega_1 |\psi^{-}\rangle\langle \psi^{-}| +\omega_2|\phi^{+}\rangle\langle \phi^{+}|+\omega_3 |\phi^{-}\rangle\langle \phi^{-}|&&\nonumber\\
    +\omega_4|\psi^{+}\rangle\langle \psi^{+}|\qquad\qquad\qquad\qquad\qquad\quad\quad&&
\end{eqnarray}
$\textmd{\small{with}}\,\omega_i$$\small{\in}$$[0,1]\,\forall i$$=$$1,2,3,4,\,\sum_{i=1}^4\omega_i$$=$$1\,\textmd{\small{and}}\,|\phi^{\pm}\rangle$$=$$\frac{|00\rangle\pm|11\rangle}{\sqrt{2}},$
$ |\psi^{\pm}\rangle$$=$$\frac{|01\rangle\pm|10\rangle}{\sqrt{2}}\,\textmd{denote the Bell states.\,Correlation matrix}$ of this class is given by: $\textmd{diag}(1 - 2 (\omega_1 + \omega_3),1 - 2 (\omega_2 + \omega_3),1 - 2 (\omega_1 + \omega_2))$. Let us consider a subclass of the family (Eq.(\ref{belldiag})) specified by $\omega_3$$=$$0.$  Members from this subclass are Bell-CHSH local if:
\begin{equation}\label{belldiag1}
    \sum_{i=1}^2(1-2\omega_i)^2\leq 1.
\end{equation}
As discussed above, nontrilocal correlations cannot be detected in any linear trilocal network involving Bell-CHSH local states. Now, let each of the sources in the
triangle network distribute an identical copy of the Bell diagonal state characterized by $\omega_3$$=$$0.$ Let the parties perform the local measurement on the basis given by Eq.(\ref{tr9}).

\begin{center}
\begin{figure}
\includegraphics[width=2.7in]{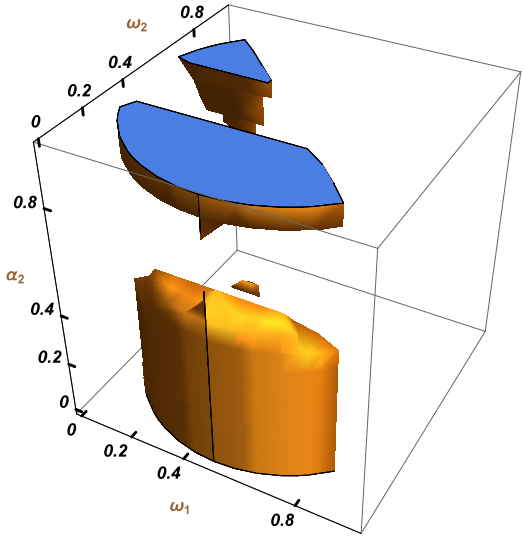} \\
 \caption{\emph{ Shaded region is a subspace of parameter space $(\omega_1,\omega_2,\alpha_2).$ Corresponding Bell diagonal states (for $\omega_3$$=$$0$)
 are Bell-CHSH local (Eq.(\ref{belldiag1})). However nontrilocal correlations are detected (using Eq.(\ref{belldiag2})) by using identical copies of these states in a triangle network for any value of the measurement parameter $\alpha_2$ lying in this region.}}
\label{fig4}
 \end{figure}
 \end{center}

Violation of the inequality (Eqs.(\ref{tr5},\ref{tr6})) occurs if:
\begin{eqnarray}\label{belldiag2}
\sqrt{|(1 - 2 \omega_2)^2 - 3\alpha_2^2 (1- 2  \omega_2)^2 + \alpha_2^4
 (2 - 6 \omega_2 + 4 \omega_2^2)|}+&&\nonumber\\
\sqrt{|(1 - 2 \omega_2)^2 (-1 - \alpha_2^2 + 4 \alpha_2^4 + 2 \omega_2)|}>1.&&\nonumber\\
 &&
\end{eqnarray}
It is observed that some members of this family (see Fig.\ref{fig4}), satisfy both the above relations (Eqs.(\ref{belldiag1},\ref{belldiag2})). Hence, the violation of Eqs.(\ref{tr5},\ref{tr6}) does not stem from the Bell-CHSH violation by the individual states involved. Consequently, the nontrilocality of
the network correlations can be detected using at least one trilocal inequality (Eq.(\ref{tr5})) irrespective of Bell's nonlocality of the states used in the triangle network.\\
\subsection{Higher Noise Tolerance}
\begin{center}
\begin{figure}
\includegraphics[width=2.7in]{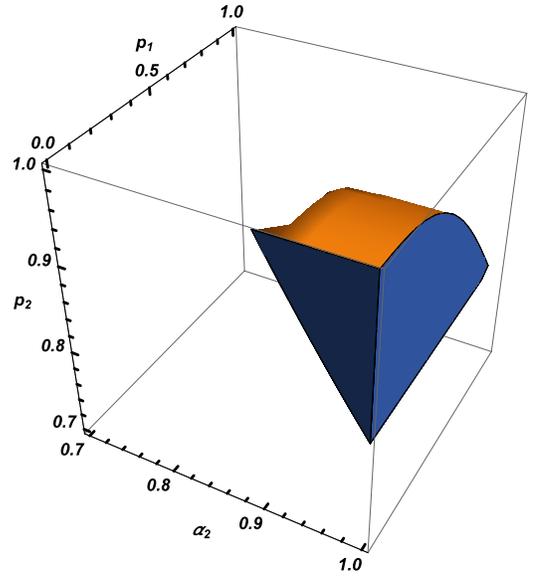} \\
 \caption{\emph{ Shaded region indicates the measurement parameter ($\alpha_2$) and noise parameters($p_1,p_2$) for which nontrilocal correlations are observed in a triangle network but cannot be detected in a linear trilocal network involving identical copies of the noisy state (Eq.(\ref{had2})).}}
\label{figsnew1}
 \end{figure}
 \end{center}

For suitable choice of $f,g,h$ there exist trilocal inequalities (Eq.(\ref{tr5})) which are more tolerant to noise compared to the existing inequality (Eq.(\ref{ineq})) for linear networks. For the sake of discussion, let us consider the particular inequality given by Eqs.(\ref{tr5},\ref{tr6}). Let three identical copies of the noisy state (Eq.(\ref{had2})) be used in a linear trilocal network. Correlation tensor of this family of states is given by $\textmd{diag}(-p_1p_2,p_1p_2,p_2).$ Using Eq.(\ref{tribd}), it is clear that nontrilocality cannot be detected in the linear network if:
\begin{equation}\label{noisy10}
    \textmd{Max}(\sqrt{2p_1^3p_2^3},p_2^{\frac{3}{2}}\sqrt{1+p_1^3})\leq 1.
\end{equation}
But when these identical copies are used in the triangle network, nontrilcality is detected if Eq.(\ref{had3}) is satisfied. For some members of this family of states (Eq.(\ref{had2})), nontrilocality is detected in the triangle network but not in a linear trilocal network (see Fig.\ref{figsnew1}).
Only one example of trilocal inequality for the triangle network has been discussed here. However, one may find further examples of trilocal inequalities for the triangle network (suitably choosing f, g and h) such that those are more resistant to noise compared to Eq.(\ref{ineq}).
\section{Polygon Network Scenario}\label{poly}
$\textmd{Consider a closed network having}\, n\,\textmd{edges and}\, n\, \textmd{vertices}.$ $\mathcal{S}_1,\mathcal{S}_2,...,\mathcal{S}_n$ denote $n$ independent sources corresponding to $n$ sides$.\,$ The network involves $n$ parties $\mathcal{P}_1,\mathcal{P}_2,....,\mathcal{P}_n$ corresponding to $n$ vertices (see Fig.\ref{fig9}). Each source $\mathcal{S}_i$ distributes particles to the parties $\mathcal{P}_i,\mathcal{P}_{i+1}\,(i$$=$$1,2,...,n-1)$ and $\mathcal{S}_n$ sends particles to parties $\mathcal{P}_n$ and $\mathcal{P}_{1}.$
\begin{center}
\begin{figure}
\includegraphics[width=3.8in]{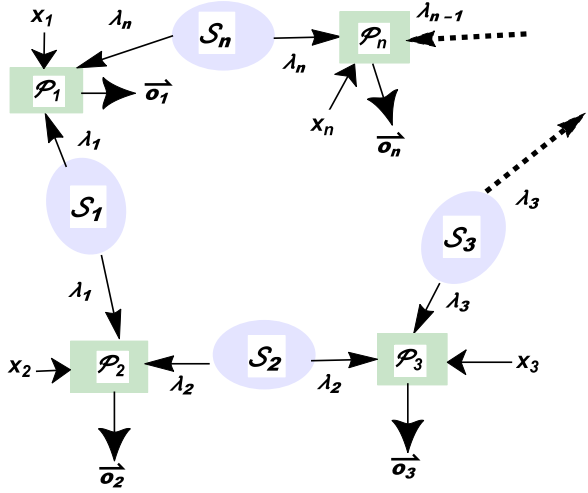} \\
 \caption{\emph{ Schematic representation of polygon network with independent sources.}}
\label{fig8}
 \end{figure}
 \end{center}
Let local variable $\lambda_i$ characterize $\mathcal{S}_i\,\forall i.$  $n$-local (source independence) assumption
takes the form:
\begin{equation}\label{trg3}
    \rho(\lambda_1,\lambda_2,...,\lambda_n)=\Pi_{i=1}^n\rho_i(\lambda_i),
\end{equation}
where $\rho_i$ corresponds to normalized stochastic distribution of $\lambda_i, \forall i.$ Let $\textsl{x}_i$ denote the fixed local ternary input of $\mathcal{P}_i\,\forall i.$ Let outcomes of $\textsl{x}_i$ be labeled as $\vec{\mathfrak{o}}_i$$=$$(\mathfrak{o}_{i1},\mathfrak{o}_{i2})$ with each component $\mathfrak{o}_{ij}$ being binary-valued. $n$-partite local correlations
 are of the form:
\begin{eqnarray}\label{trg4}
&&p(\vec{\mathfrak{o}}_1,...,\vec{\mathfrak{o}}_n)=
\int_{\Lambda_1}...\int_{\Lambda_n} d\lambda_1...d\lambda_n \rho(\lambda_1,...,\lambda_n)R_3\\
&&\textmd{where}\,R_3=\Pi_{i=2}^{n}p(\vec{\mathfrak{o}}_i|\lambda_i,\lambda_{i-1})
     p(\vec{\mathfrak{o}}_1|\lambda_1,\lambda_n).\nonumber
\end{eqnarray}
Any set of $n$-partite correlations generated in the network is $n$-local if it is local, i.e., satisfies Eq.(\ref{trg4}) and also satisfies the $n$-local
constraint (Eq.(\ref{trg3})). Correlation inexplicable in this form (Eqs.(\ref{trg3},\ref{trg4})) is said to be \textit{non $n$-local}.\\

The set of $n$-local inequalities characterizing correlations is given by the following theorem.\\
\textit{Theorem.2:}\,In a polygon network, an arbitrary set of $n$-local correlations necessarily satisfies:
\begin{equation}\label{trg5}
    \sqrt{|I_{1,n}|}+\sqrt{|I_{2,n}|}\leq 1,
\end{equation}
\textmd{where} $\forall j=1,2\,\,\textmd{and for every fixed}\,t$$=$$1,...,n,$
\begin{eqnarray}\label{trg6i}
  I_{j,n} = \sum_{\mathcal{C}_1}(-1)^{j\sum_{s\in \mathcal{C}_2}i_s}\langle \Pi_{s\in \mathcal{C}_2}C_{s}^{i_s} C_{t}^{j-1}\rangle,\,\textmd{\small{where}}\qquad\qquad\qquad&&\nonumber\\
\langle \Pi_{s\in \mathcal{C}_2}C_{s}^{i_s} C_{t}^{j-1}\rangle =\sum_{\mathcal{C}_3}(-1)^{\sum_{s\in \mathcal{C}_2}f_s(i_s,\vec{\mathfrak{o}_s})+
  f_t(j-1,\vec{\mathfrak{o}_t})}p(\vec{\mathfrak{o}}_1,...,\vec{\mathfrak{o}}_n)&&\nonumber\\
   \textmd{with}\,\,\mathcal{C}_1=\{i_k|k\neq t\}, \,\mathcal{C}_2=\{k=1,2,...,n|k\neq t\},\qquad\quad\quad&&\nonumber\\
   i_k\in\{0,1\}\,\textmd{and}\,\mathcal{C}_3=\{\mathfrak{o}_{11},\mathfrak{o}_{12},\mathfrak{o}_{21},\mathfrak{o}_{22},....,
   \mathfrak{o}_{n1},\mathfrak{o}_{n2}\}.\qquad\qquad&&\nonumber
\end{eqnarray}
In above expression, $\forall i,\,f_i(.,.)$ denotes arbitrary integer-valued function.


\textit{Proof:}\,Proof is based on the same procedure that was used for proving the previous theorem.\\

The role of the parties is interchangeable in the correlator terms appearing in Eq.(\ref{trg5}). The above theorem characterizes any $n$-local correlation arising in a $n$-sided closed network. Violation of at least one of the inequalities specified by Eq.(\ref{trg5}) detects non $n$-locality in the network. However, such a detection criterion being only sufficient, there may exist non $n$-local correlations satisfying Eq.(\ref{trg5}) for all possible forms of the integer-valued functions. \\
For a demonstration of quantum violation of Eq.(\ref{trg5}), let us consider that each of $n$ sources generates a two-qubit state. On receiving the qubits, let each of the parties measure on the basis given by Eq.(\ref{tr9}). It is conjectured that for suitable choice of the functions $f_1,...,f_n,$ Eq.(\ref{trg5}) will be violated. An instance of violation is provided for $n$$=4.$\\
Let each of $\mathcal{S}_1,\mathcal{S}_2,\mathcal{S}_3,\mathcal{S}_4$ generate a Bell state (Eq.(\ref{tr8})) in a square network. Consider a particular $4$-local inequality given by Eq.(\ref{trg5}) for $t$$=$$2$ and integer-valued functions specified as follows:
\begin{eqnarray}\label{trg6iv}
   f_1(.,.)= f_4(.,.)=h(.,.)\,\textmd{\small{and}}\,f_2(.,.)=f_3= f(.,.\,)&&\\
   \textmd{where functions}\,f,h\,\textmd{are given by Eq.(\ref{tr6iv})}\qquad&&\nonumber
\end{eqnarray}
Correlations violate the $4$-local inequality (Eqs.(\ref{trg5},\ref{trg6iv})) if:
\begin{equation}\label{4loc}
     \sqrt{|2 - 5 \alpha_2^2 + 8\alpha_2^4|}+\sqrt{|\alpha_2^2 (3 - 2 \alpha_2^2)|}>2^\frac{3}{2}.
\end{equation}
Violation is observed over some range of $\alpha_2$ (see Fig.\ref{fig9}).\\
\begin{center}
\begin{figure}
\includegraphics[width=2.7in]{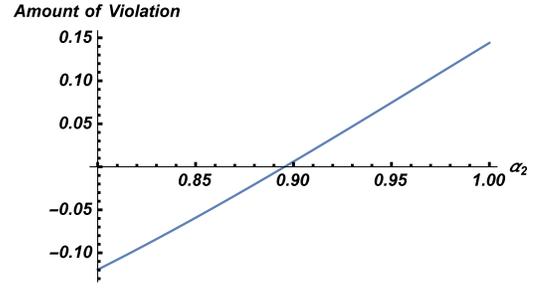} \\
 \caption{\emph{ Plotting range of measurement parameter $\alpha_2$ for which $4$-local inequality (Eqs.(\ref{trg5},\ref{trg6iv})) is violated, i.e., the relation given by Eq.(\ref{4loc}) is  satisfied.}}
\label{fig9}
\end{figure}
\end{center}
\section{Discussions}\label{conc}
Measurement independence is one of the potent factors in manifesting nonlocality in standard Bell scenario\cite{brunrev}. However, there exist examples\cite{gis1} of nonlocality (nontrilocality) in a network scenario where each of the parties performs a fixed measurement. The present work deals with the study of such correlations introducing trilocal inequalities for any triangle network followed by an extension to a polygon network. More specifically, a complete set of Bell-type inequalities in a triangle network is provided under the source independence assumption. Violation of at least one such inequality is sufficient to detect nontrilocality in the network. As already discussed, these inequalities, unlike the existing trilocal inequality (compatible with linear trilocal networks\cite{km1,bilo5}) can detect nontrilocality in a network using Bell-CHSH local states. To some extent, such an observation is similar to the demonstration of tripartite entanglement of a state even when bipartite entanglement of none of the reduced states can be detected. It will be interesting to further exploit such an analogy between multipartite entanglement detection and non $n$-locality detection in closed quantum networks.\\
For analysis of quantum violation, a few specific local measurement bases (Eqs.(\ref{tr9},\ref{tr9new})) have been considered. Consideration of more general measurement contexts may contribute toward a better characterization of network correlations. \\
The entire analysis is confined to closed networks where each source distributes particles to only two parties. It will be interesting to study networks where each source generates a $m$-partite ($m$$\geq$$3$) state. Framing Bell-type inequalities for such a scenario is important in the perspective of developing an experimentally feasible genuine entanglement detection scheme.\\
To that end, it may be noted here that the correlations being manifested in fixed input scenario, the inequalities are devoid of freedom-of-choice loophole\cite{lp3}. However, the same may suffer from several other loopholes in experimental scenarios. In this context, analysis of noise tolerance of some of these detection criteria is provided for some specific cases. However, a more general treatment of the same can be considered as a direction of potential future research.
\section*{Acknowledgement}
I am thankful to the anonymous referees of Physical Review A journal for thoroughly checking the manuscript and providing valuable suggestions which helped in improving the work. Also, I would like to thank my brother Mr.Puranjan Mukhopadhyay for helping me to improve my writing style and grammar.
\section*{Appendix.A}\label{appa}
\textit{Proof of the theorem:} By assumption of the theorem, the correlations are trilocal (Eq.(\ref{tr4})). Let us define the following terms:
\begin{eqnarray}\label{app1}
 \langle C_1^i\rangle_{\lambda_1,\lambda_3}&=&\sum_{\mathfrak{o}_{11},\mathfrak{o}_{12}}(-1)^{f(i,\vec{\mathfrak{o}_1})}p_{\lambda_1,\lambda_3}(\vec{\mathfrak{o}}_1)\nonumber\\
   \langle C_2^i\rangle_{\lambda_1,\lambda_2}&=&\sum_{\mathfrak{o}_{21},\mathfrak{o}_{22}}(-1)^{g(i,\vec{\mathfrak{o}_2})}p_{\lambda_1,\lambda_2}(\vec{\mathfrak{o}}_2)\nonumber\\
  \langle C_3^i\rangle_{\lambda_2,\lambda_3}&=&\sum_{\mathfrak{o}_{31},\mathfrak{o}_{32}}(-1)^{h(i,\vec{\mathfrak{o}_3})}p_{\lambda_2,\lambda_3}(\vec{\mathfrak{o}}_3)\,\,\textmd{with}\, i=0,1\nonumber\\
  \end{eqnarray}

Functions $f,g,h$ are used in framing the terms $\langle C_1^iC_2^{j-1}C_3^k\rangle$ in Eq.(\ref{tr6i}). Each of the three functions appears as an exponent of $-1.$ So, for any even value of these functions, $+1$ is obtained whereas, for any odd value of these functions, $-1$ is obtained. So, each probability term appearing in the summation in the expression of any of the marginal expectations Eq.(\ref{app1}), is having either $+1$ or $-1$ as a coefficient. Again, adding up all probability terms appearing in the expression of any of these terms gives $1.$ So, combining these two facts, for any fixed value of hidden variables $\lambda_1,\lambda_2,\lambda_3$, absolute value of any of these terms (Eq.(\ref{app1})) is less than $1.$ Hence,
\begin{equation}\label{absl}
   |\langle C_1^i\rangle_{\lambda_1,\lambda_3}|, |\langle C_2^i\rangle_{\lambda_1,\lambda_2}|,|\langle C_3^i\rangle_{\lambda_2,\lambda_3}|\leq 1,\,i=0,1
\end{equation}
Now consider any $(\lambda_1,\lambda_2,\lambda_3)$$\in$$\Lambda_1\otimes\Lambda_2\otimes\Lambda_3.$\\
$\forall\, i,j,k$$=$$0,1,$ we have,
\begin{widetext}
\begin{eqnarray}\label{ap1}
 \langle C_1^i C_2^{j}C_3^k\rangle_{\lambda_1,\lambda_2,\lambda_3} &=& \sum_{\mathcal{C}}(-1)^{f(i,\vec{\mathfrak{o}_1})+
   g(j,\vec{\mathfrak{o}_2})+h(k,\vec{\mathfrak{o}_3})}p_{\lambda_1,\lambda_2,\lambda_3}(\vec{\mathfrak{o}}_1,\vec{\mathfrak{o}}_2,\vec{\mathfrak{o}}_3)\,\,
   \textmd{with}\,\mathcal{C}=\{\mathfrak{o}_{11},\mathfrak{o}_{12},\mathfrak{o}_{21},\mathfrak{o}_{22},
   \mathfrak{o}_{31},\mathfrak{o}_{32}\}\nonumber\\
 &=&\sum_{\mathcal{C}}(-1)^{f(i,\vec{\mathfrak{o}_1})+
   g(j,\vec{\mathfrak{o}_2})+h(k,\vec{\mathfrak{o}_3})}p_{\lambda_1,\lambda_3}(\vec{\mathfrak{o}}_1).p_{\lambda_1,\lambda_2}(\vec{\mathfrak{o}}_2)
   .p_{\lambda_2,\lambda_3}(\vec{\mathfrak{o}}_3) \nonumber \\
    &=&(\sum_{\mathfrak{o}_{11},\mathfrak{o}_{12}}(-1)^{f(i,\vec{\mathfrak{o}_1})}p_{\lambda_1,\lambda_3}(\vec{\mathfrak{o}}_1))
   (\sum_{\mathfrak{o}_{21},\mathfrak{o}_{22}}(-1)^{g(j,\vec{\mathfrak{o}_2})}p_{\lambda_1,\lambda_2}(\vec{\mathfrak{o}}_2))
   (\sum_{\mathfrak{o}_{31},\mathfrak{o}_{32}}(-1)^{h(k,\vec{\mathfrak{o}_3})}p_{\lambda_2,\lambda_3}(\vec{\mathfrak{o}}_3))\nonumber\\
    &=& \langle C_1^i\rangle_{\lambda_1,\lambda_3}\langle C_2^{j}\rangle_{\lambda_1,\lambda_2}\langle C_3^k\rangle_{\lambda_2,\lambda_3}
 \end{eqnarray}
 Hence,
\begin{eqnarray}\label{ap2}
(I_1)_{\lambda_1,\lambda_2,\lambda_3}=\langle C_1^0-C_1^1\rangle_{\lambda_1,\lambda_3}\langle C_2^0\rangle_{\lambda_1,\lambda_2}
\langle C_3^0-C_3^1\rangle_{\lambda_2,\lambda_3}\nonumber\\
(I_2)_{\lambda_1,\lambda_2,\lambda_3}=\langle C_1^0+C_1^1\rangle_{\lambda_1,\lambda_3}\langle C_2^1\rangle_{\lambda_1,\lambda_2}
\langle C_3^0+C_3^1\rangle_{\lambda_2,\lambda_3}
\end{eqnarray}
Now,

\begin{eqnarray}\label{app2}
  I_1 &=& \frac{1}{4}\int_{\Lambda_1}\int_{\Lambda_2}\int_{\Lambda_3}\Pi_{i=1}^3\rho_i(\lambda_i)d\lambda_i(I_1)_{\lambda_1,\lambda_2,\lambda_3}\,\,
  \textmd{(By the trilocality constraint Eq.(\ref{tr3}))}  \nonumber\\
   &=& \frac{1}{4}\int_{\Lambda_1}\int_{\Lambda_2}\int_{\Lambda_3}\Pi_{i=1}^3\rho_i(\lambda_i)d\lambda_i(\langle C_1^0-C_1^1\rangle_{\lambda_1,\lambda_3})
  \langle C_2^0\rangle_{\lambda_1,\lambda_2}(\langle C_3^0-C_3^1\rangle_{\lambda_2,\lambda_3})\,\,(\textmd{By Eq.(\ref{ap2})})\nonumber \\
     &=&\frac{1}{4} \int_{\Lambda_1}\int_{\Lambda_2}\Pi_{i=1}^2\rho_i(\lambda_i)\langle C_2^0\rangle_{\lambda_1,\lambda_2}d\lambda_i
     \int_{\Lambda_3}\rho_3(\lambda_3)d\lambda_3( \langle C_1^0-C_1^1\rangle_{\lambda_1,\lambda_3})
( \langle C_3^0-C_3^1\rangle_{\lambda_2,\lambda_3})
 \end{eqnarray}
Now for any fixed $\lambda_1$$\in$$\Lambda_1$ and $\lambda_2$$\in$$\Lambda_2,$
\begin{eqnarray}\label{app3}
  \int_{\Lambda_3}\rho_3(\lambda_3) \langle C_1^0-C_1^1\rangle_{\lambda_1,\lambda_3}
 \langle C_3^0-C_3^1\rangle_{\lambda_2,\lambda_3}d\lambda_3&=& \int_{\Lambda_3}\rho_3(\lambda_3)(\sum_{\mathfrak{o}_{11},\mathfrak{o}_{12}}(
 (-1)^{f(0,\vec{\mathfrak{o}_1})} p_{\lambda_1,\lambda_3}(\vec{\mathfrak{o}}_1)-(-1)^{f(1,\vec{\mathfrak{o}_1})} p_{\lambda_1,\lambda_3}(\vec{\mathfrak{o}}_1)))\nonumber\\
 && (\sum_{\mathfrak{o}_{31},\mathfrak{o}_{32}}(
 (-1)^{h(0,\vec{\mathfrak{o}_3})} p_{\lambda_2,\lambda_3}(\vec{\mathfrak{o}}_3)-(-1)^{h(1,\vec{\mathfrak{o}_3})} p_{\lambda_2,\lambda_3}(\vec{\mathfrak{o}}_3)))
  d\lambda_3\nonumber\\
   &=&(\sum_{\mathfrak{o}_{11},\mathfrak{o}_{12}}(
 (-1)^{f(0,\vec{\mathfrak{o}_1})} p_{\lambda_1}(\vec{\mathfrak{o}}_1)-(-1)^{f(1,\vec{\mathfrak{o}_1})} p_{\lambda_1}(\vec{\mathfrak{o}}_1)))\nonumber\\
 && (\sum_{\mathfrak{o}_{31},\mathfrak{o}_{32}}(
 (-1)^{h(0,\vec{\mathfrak{o}_3})} p_{\lambda_2}(\vec{\mathfrak{o}}_3)-(-1)^{h(1,\vec{\mathfrak{o}_3})} p_{\lambda_2}(\vec{\mathfrak{o}}_3)))
  \nonumber\\
  &&\textmd{where}\, p_{\lambda_i}(\vec{\mathfrak{o}_j})=\int_{\Lambda_3}\rho_3(\lambda_3) p_{\lambda_i,\lambda_3}(\vec{\mathfrak{o}_j})d\lambda_3,\,\forall
  i=1,2;\,\,j=1,3\nonumber\\
   &=& \langle C_1^0-C_1^1\rangle_{\lambda_1}
 \langle C_3^0-C_3^1\rangle_{\lambda_2}
\end{eqnarray}
Using Eq.(\ref{app3}), we get from Eq.(\ref{app2}),
\begin{eqnarray}\label{app4}
  |I_1| &=& \frac{1}{4}|\int_{\Lambda_1}\int_{\Lambda_2}\Pi_{i=1}^2\rho_i(\lambda_i)d\lambda_i(\langle C_1^0-C_1^1\rangle_{\lambda_1})
  \langle C_2^0\rangle_{\lambda_1,\lambda_2}(\langle C_3^0-C_3^1\rangle_{\lambda_2})|\nonumber \\
     &\leq&\frac{1}{4}\int_{\Lambda_1}\int_{\Lambda_2}\Pi_{i=1}^2\rho_i(\lambda_i)d\lambda_i|\langle C_1^0-C_1^1\rangle_{\lambda_1}|
  |\langle C_2^0\rangle_{\lambda_1,\lambda_2}||\langle C_3^0-C_3^1\rangle_{\lambda_2}|\nonumber \\
  &\leq&\frac{1}{4} \int_{\Lambda_1}\int_{\Lambda_2}\Pi_{i=1}^2\rho_i(\lambda_i)d\lambda_i|\langle C_1^0-C_1^1\rangle_{\lambda_1}|
  |\langle C_3^0-C_3^1\rangle_{\lambda_2}|,\,\textmd{(By Eq.(\ref{absl}))}\nonumber \\
  &=&\frac{1}{4}( \int_{\Lambda_1}\rho_1(\lambda_1)d\lambda_1|\langle C_1^0-C_1^1\rangle_{\lambda_1}|)(
  \int_{\Lambda_2}\rho_2(\lambda_2)d\lambda_2|\langle C_3^0-C_3^1\rangle_{\lambda_2}|)
 \end{eqnarray}
Similarly it can be shown that the other correlator term $I_2$ satisfies:
\begin{equation}\label{app5}
    |I_2|\leq \frac{1}{4}( \int_{\Lambda_1}\rho_1(\lambda_1)d\lambda_1|\langle C_1^0+C_1^1\rangle_{\lambda_1}|)(
  \int_{\Lambda_2}\rho_2(\lambda_2)d\lambda_2|\langle C_3^0+C_3^1\rangle_{\lambda_2}|)
\end{equation}
Using the relation given by Eq.(\ref{prodbas7}) we get,
\begin{equation}\label{app6}
    \sqrt{|I_1|}+    \sqrt{|I_2|}\leq ( \int_{\Lambda_1}\rho_1(\lambda_1)d\lambda_1\frac{|\langle C_1^0+C_1^1\rangle_{\lambda_1}|
    +|\langle C_1^0-C_1^1\rangle_{\lambda_1}|}{2})^{\frac{1}{2}}(
  \int_{\Lambda_2}\rho_2(\lambda_2)d\lambda_2\frac{|\langle C_3^0+C_3^1\rangle_{\lambda_2}|
    +|\langle C_3^0-C_3^1\rangle_{\lambda_2}|}{2})^{\frac{1}{2}}
\end{equation}
Now $\frac{1}{2}(|\langle C_1^0\rangle_{\lambda_1}+\langle C_1^1\rangle_{\lambda_1}|+|\langle C_1^0\rangle_{\lambda_1}-\langle C_1^1\rangle_{\lambda_1}|)
$ $ =$ $\small{\textmd{max}}\{|\langle C_1^0\rangle_{\lambda_1}|,|\langle C^1_1\rangle_{\lambda_1}|\}\leq 1.$ Similarly for observables $C_3^0$ and $C_3^1.$
Eq.(\ref{app6}) thus gives:
\begin{eqnarray}
  \sqrt{|I_1|}+    \sqrt{|I_2|}&\leq&  (\int_{\Lambda_1}\rho_1(\lambda_1)d\lambda_1)^{\frac{1}{2}}(\int_{\Lambda_2}\rho_2(\lambda_2)d\lambda_2)^{\frac{1}{2}}\nonumber\\
  &=& 1
\end{eqnarray}
\end{widetext}
$$\,$$
\section*{Appendix.b}\label{appb}
Here we present the expression of the trilocal inequality given by Eqs.(\ref{tr5},\ref{tr6}) in the network involving two pure entangled states(Eq.(\ref{ent1})) and one pure product state (Eq.(\ref{prodbas4})). Depending on which of the two sources generate entangled states, the expressions differ as follows:
\begin{widetext}
\begin{center}
\begin{table}[htp]
\caption{}
\begin{center}
\begin{tabular}{|c|c|}
\hline
States&Expression($\mathcal{T}$)\\
\hline
&\\
$\varrho_{ent}^{1},\varrho_{ent}^{2}$&$\frac{1}{\sqrt{2}}(\sqrt{|\small{(\tau_{11}^2) \tau_{22}^2 (1 - \alpha_2^2 (1 + u_{53})) (1 -
   \alpha_2^2 (1 - u_{63}) + u_{63})}|}$\\
  $\varrho_{prod}^{(5,6)}$ &$ +\sqrt{|u_{53} ((-1 + 3 \alpha_2^2 - 2 \alpha_2^4) \tau_{21}^2 \tau_{22}^2 (-1 +
      u_{63}) + (1 - 2 \tau_{21}^2 +
      2 \alpha_2^2 (-1 + \tau_{21}^2)) \tau_{12}^2 (-1 + u_{63} +
      \alpha_2^2 (1 + u_{63})))|})$\\
      &\\
      \hline
      &\\
$\varrho_{ent}^{2},\varrho_{ent}^{3}$&$\frac{1}{\sqrt{2}}(\sqrt{|\small{\tau_{22}^2 (-\alpha_2^4 \tau_{23}^2 (-1 + u_{1,3}) - \tau_{12}^2 u_{1,3} +
   \alpha_2^2 (\tau_{23}^2 u_{1,3} + \tau_{12}^2 (1 + u_{1,3}))) (1 + u_{2,3})}}+$\\
  $\varrho_{prod}^{(1,2)}$ &$ \sqrt{|\small{u_{1,3} (\tau_{23}^2 (2 - \tau_{22}^2 (1 + u_{2,3})) +
   2 \alpha_2^4 (\tau_{22}^2 \tau_{23}^2 (1 - u_{2,3}) - \tau_{12}^2 \tau_{13}^2 (1 + u_{2,3}))-
   \alpha_2^2 (2 (-\tau_{12}^2) \tau_{13}^2 + \tau_{23}^2 (2 + \tau_{22}^2 (1 - 5 u_{2,3}) + 2 u_{2,3})))}|}$\\

&\\
\hline
&\\
$\varrho_{ent}^{1},\varrho_{ent}^{3}$&$\frac{1}{\sqrt{2}}(\sqrt{|\small{(-\tau_{11}^2) (2 \tau_{13}^2 u_{3,3} +
   2 \alpha_2^4 \tau_{13}^2 (1 + u_{3,3}) -
   \alpha_2^2 (\tau_{23}^2 (-1 + u_{3,3}) + \tau_{13}^2 (1 + 5 u_{3,3}))) (-1 +
   u_{4,3})}|}$\\

  $\varrho_{prod}^{(3,4)}$ &$ +\sqrt{|(2 \tau_{23}^2 u_{3,3} +
   \alpha_2^2 (-\tau_{23}^2 (-1 + u_{3,3}) + \tau_{13}^2 (1 +
         u_{3,3}))) (1 + \tau_{21}^2 (-1 - 3 u_{4,3}) + u_{4,3} +
   \alpha_2^2 (-2 + (-2 + 4 \tau_{21}^2) u_{4,3}))|})$\\
   &\\
\hline
\end{tabular}
\end{center}
\label{table:ta1}
\end{table}
\end{center}
\end{widetext}

 \end{document}